\pdfoutput=1
\documentclass[10pt]{article}

\usepackage{algorithm}
\usepackage{color}
\usepackage{bbm}
\usepackage{epsfig}
\usepackage{amssymb,amsmath,amscd, mdwmath}
\usepackage{graphicx,cite,url}
\usepackage{algorithmic}
\usepackage{float}
\usepackage{amsfonts}
\usepackage{syntonly}
\usepackage{multirow}
\usepackage{graphicx}
\usepackage{subfig}
\usepackage{tabularx}
\usepackage{setspace}
\usepackage{stfloats}
\usepackage{amsthm}
\usepackage[labelfont=bf,labelsep=period,justification=raggedright]{caption}

\makeatletter
\renewcommand{\@biblabel}[1]{\quad#1.}
\makeatother

\date{}

\begin{document}

\begin{flushleft}
{\Large
\textbf{A distributed hypergraph model  for  the full-scale simulation of the collaborations in dblp }
}
\\
Zheng Xie$^{1, \sharp }$
\\
\bf{1}   College of Liberal Arts and Sciences,   National University of Defense Technology, Changsha,   China.
\\  $^\sharp$ xiezheng81@nudt.edu.cn
 \end{flushleft}
\section*{Abstract}
This study proposed  a   model to give a    full-scale simulation for  the dynamics of  the  collaborations    in the dblp dataset.
It is a distributed    model with the capability  of    simulating   large  hypergraphs, namely   systems with heterogeneously multinary relationship.
 Its assembly mechanism of hyperedges is driven by   Lotka's law and a cooperative game that maximizes   benefit-cost ratio for collaborations.
The model
is built on a circle to express the game, expressing the cost
by the distance between nodes. The
benefit of coauthoring with a productive researcher or one with many coauthors is  expressed by
the cumulative degree or hyperdegree of nodes.
The model successfully
captures  the multimodality of collaboration patterns  emerged in the  dblp dataset, and reproduces the evolutionary trends  of collaboration pattern, degree, hyperdegree, clustering, and giant component
over  thirty years remarkably well.
This  model has the potential to be extended to
  understand the complexity of     self-organized systems
that  evolve mainly  driven by
   specific  cooperative games, and would be capable of
   predicting the behavior patterns  of  system  nodes.






%


\noindent {\bf Keywords:}  Coauthorship network,  Multimodality, Data modelling.

\section*{Introduction}

A growing trend of collaboration has  emerged in    current scientific research, which  is  reflected by the  increasingly active coauthorship among researchers as solitary authorship diminishes in prevalence\cite{Committee}.
Coauthorship is  a manifestation of team work on research.
 A  field known as  team science  draws on diverse disciplinary perspectives to understand  the process  and outcomes of scientific collaborations.
  The assembly patterns of
  research teams  determine the  structure of coauthorship networks\cite{Newman2004patterns}, and connect  with
academic  performances, such as   citation impacts\cite{Wuchty-Jones2007,Valderas2007}, transdisciplinary outcomes\cite{Vogel2014}, and   publication productivity\cite{Hall2012}.
 The teams with a higher fraction of incumbents, who contribute expertise and knowledge
to the teams, have a higher probability to publish papers on the journals with higher impact factor,
whereas the  teams with a preponderance of repeating past collaborations  have
a lower  probability\cite{Guimera-Uzzi2005}.
  Therefore, it is meaningful  to build models for the simulation and
prediction  of  collaboration patterns.

Modelling the collaborations  among researchers also helps us elucidate   the important question ``how did cooperative behavior evolve\cite{Pennisi2005}?"
It needs to explore the mechanisms underlying collaborations, including the mechanisms of authors   assembling to write  papers and    the mechanisms of authors joining  scientific communities. This exploration  helps us understand   the evolution, complexity, and multimodality of scientific   collaborations.
Previous  studies on modelling  collaborations
 concentrate on  the   distribution of the number of coauthors\cite{BarabasiJeong2002}, followed by  network structure \cite{Newman1, Zhang-Bu2018}, and phase transition\cite{XieO2016,XieO2018}. To reproduce them,
researchers  have a range of models to try possible mechanisms, from  preference attachment to cooperative game theory\cite{Santos,Xie2018Complexity}.
Most of these models generate a constant number of links for each new   node,  far from the reality, and cannot give  simulations in full scale.


The dblp computer science bibliography provides a high-quality dataset  that consists of   open bibliographic information on the major journals and conference proceedings of computer science\footnote[1]{https://www.dblp.org}.
 It has been corrected by several methods of name disambiguation, and there are now more than 60,000 manually
confirmed external identities linked with dblp author bibliographies.
 We proposed a model to give  a full-scale simulation for the collaborations in the dblp dataset  from $1986$ to $2015$, which involves 149,285  authors  and their collaborations for 106,821 papers in 1,304 journals and conference proceedings.
The system of researchers and their collaborations   evolve in a parallel mode,   researchers and   teams   publishing  papers concurrently. Therefore,
 the proposed   model  is designed in a distributed way.

 Our distributed model is based on  Lotka's law and a cooperative game of maximizing  the benefit-cost ratio of  collaboration.
To express the game, we built the model as a geometric hypergraph on a circle, used
the distance between nodes to express the cost, and used the cumulative degree and hyperdegree to express
the benefit of coauthoring with a  researcher. The number of
  hyperedges and their size are the same to those of the empirical  dataset. The number of new  nodes at each time step is proportional to the number of the authors newly appearing in the empirical dataset at that time.
 Experiments show that the model successfully
captures a range of characteristics of  the dblp  dataset    in full scale, such as the evolutionary trends of collaboration pattern, the number of papers, the number of coauthors,
  clustering,   and giant component size  over   30 years.

This paper is organized as follows. Literature review and empirical data are described in Sections 2,
3. The model is described in Sections 4, 5, where its mechanisms  are
 analyzed. The results are discussed in Section 6, and conclusions drawn in Section 7.

\section*{Literature review}

Collaboration patterns have attracted much attention, with analyses of perspectives ranging from contribution\cite{Correa2017,Lu-Zhang2019}, individual\cite{Glanzel2014},  population\cite{Li-Miao2015},      discipline\cite{Moody2004,Wagner,XieLL}, country\cite{Perc,Katz1994}, multination\cite{Russell1995,Glanzel-Schubert1999,Leclerc-Gagne1994,Gomez-Fernandez1999}, and the
 relationship with citations\cite{Narin-Stevens1991,Khor-Yu2016,XieX2017}.
  The system of collaborations  can be expressed   as  a hypergraph, in which  a node
 represents an author  and  a hyperedge    represents the coauthorship in a paper. The  size of a hyperedge is  the number    of the authors of a paper.
 The simple graph extracted from that hypergraph is termed   coauthorship network\cite{Glanzel-Schubert2004}, where edges are generated by connecting each pair of the nodes belonging to the same hyperedge.
Coauthorship networks attract  much attention of researchers in social dynamics and complex science\cite{Mali-Kronegger2012,ZengA2017}.


Coauthorship networks are  featured by specific local and global features, such as  degree assortativity, high clustering, the fat tails of degree and hyperdegree distributions, small-world\cite{Newman2}.
 The degree   of an author refers  the number of his or her  coauthors,
 and the  hyperdegree of an author  refers the number of  his or her  papers.
Degree assortativity  is a preference of   nodes  tending   to connect other nodes with similar degree\cite{Newman2002assort}, and
high clustering   is     tending to cluster together\cite{Newman2001clust}. Small-world refers   a network with high clustering and its average shortest path
length scaling as the logarithm of the number of   nodes\cite{Newman3}.
 A fat-tailed distribution is a probability distribution that exhibits a large skewness, relative to that of   a normal distribution.

Researchers  explore possible  mechanisms for the evolution of those  networks. Newman et al found that the probability of a researcher coauthoring with a new one  increases with the number of his or her past coauthors,
  and   the probability of authors collaborating on writing papers  increases with the number of their common coauthors\cite{Newman2001clust}.
     The  connection  mechanisms    designed   based on the first finding      can predict high clustering.
       The  connection  mechanisms designed  based on the       second   finding, called the Matthew effect, preferential attachment, or cumulative advantage, can predict the  fat tail  of the distribution of the number of coauthors~\cite{PercM2014}.



The preferential attachment has been combined with other mechanisms to capture more features of coauthorship networks, such as   capturing degree assortativity by connecting two unconnected nodes that have similar degrees\cite{Catanzaro}, and capturing small-world by combining the mechanisms of  small-world models\cite{Ferligoj2015}.
 Barab\'asi et la proposed a  model for coauthorship networks, which connects two existing nodes with a probability proportional to the multiplication of their degrees. The model   can capture    node clustering, but cannot predict   degree assortativity\cite{BarabasiJeong2002}.
These models simulate coauthorship as binary relationship, namely connecting two nodes at a time; thus  they directly generate  graphs to discuss   the network features
of collaboration behaviors, ignoring the characters of multinary relationship in collaborations.

The number of authors of a paper can be larger than two.
Therefore, the nature of a coauthorship network is   a hypergraph. We can assemble  any number of nodes as a hyperedge to express the coauthorship in a paper.
 There is another kind of models of coauthorship networks, namely hypergraph models. For example,
 B\"orner et al proposed a model for citation and collaboration behaviors, which takes
  into consideration the effect of research topics\cite{Borner2004}.  In their model, nodes are randomly assigned with a topic, and coauthorship is modelled by randomly partitioning the  nodes with the same topic into certain groups.


 Guimer\'a et al proposed   a hypergraph model, which is controlled by  team size, the proportion of newcomers in new
hyperedges, and the proportion of incumbents to repeat previous collaborations\cite{Guimera-Uzzi2005}.
Their model starts  with an endless pool
of newcomers. Newcomers become incumbents the first time step after being selected
for a team. At each time step,
a new hyperedge $e$ is assembled and added   to the hypergraph by selecting $m$   nodes sequentially.
Each node $i$ in $e$ has a probability $p$  drawn from   incumbents and
a probability  $1-p$   drawn from   newcomers. If $i$ is drawn from
  incumbents  and there is already
another incumbent   in $e$, then $i$  has a
probability $q$ is randomly selected from  the neighbors of
a randomly selected node already in $e$, or a probability $1-q$ randomly selected
from all incumbents. Nodes that remain inactive for
longer than $\tau$ time steps are removed.

Coauthorship is a manifestation  of  the cooperations among authors. The  five typical rules in the evolution of   cooperation\cite{Nowak}  also  exist in the evolution of scientific collaborations. Coauthoring  often occurs in a research group between tutors and students, which is a  kin selection\cite{Ponomariov2016}. Cooperation contributes to academic outcomes, which is a direct reciprocity\cite{DuctorL2015}. High quality papers bring their authors   reputation, which is an indirect reciprocity. In network situation, the effect of reputation  contributes to attracting collaborators, which  is  a network reciprocity.
Cohesive research teams are easier  to attract collaborators than discordant teams, which is a group selection. Therefore,
cooperative game models have the potential to reveal the complexity emerged in coauthorship networks.
 Cooperative game for coauthoring   can be expressed by
a geometric hypergraph, where the cost  of cooperation  can be  modelled by spatial distance, and   benefit   by node hyperdegree. The cooperation condition of positive benefit-minus-cost
 can derive  the  multimodality   phenomena of coauthorship networks  in degree distribution, clustering and degree assortativity\cite{Xie2019Scientometrics}.

To sum up, the aforementioned models
 generate networks growing  from one or several nodes to large  networks with  the sizes that can be compared with empirical networks.
  The compared    empirical networks have already    grown to  certain sizes,  but the growing process   is not compared.
 There is no result  on the full-scale simulation
 for the growing process of collaborations  at a given time interval.
%



\section*{The data}

Extract two sets  from  the dblp dataset, denoted by Set 1 and Set 2, which are     at two adjacent time intervals.  Note that   the papers with more than 80 authors have been  filtered. Denote
the time interval  of Set 1 by $[T_0,T_1]$ and that of Set 2 by $(T_1,T_2]$, where the unit of time is year.
Set 1 is used to extract nodes' historical degree and hyperdegree. The collaborations in Set 2 are what we want to simulate in full scale.
In this study, $T_0=1951$, $T_1=1985$, and   $T_2=2015$.
 Table~\ref{tab1} shows some statistical indexes of the two sets.
 The proposed model will give  a full-scale simulation of the collaborations
 at the time interval $[1986, 2015]$ for the researchers who   published papers  at $[1951, 2015]$.




\begin{table}[!ht] \centering \caption{{\bf The information of test  datasets.} }
\begin{tabular}{l ccccccccccc} \hline
Dataset&  $a$   & $b$ &  $c$  & $d$ & $e$ & $f$    \\ \hline
Set 1 &1951--1985 & 5,099&  6,285& 132& 1.592&1.292\\
 Set  2   &1986--2015 &   148,928&  106,821   &1,304& 1.610 & 2.245 \\
\hline
 \end{tabular}
  \begin{flushleft} The index  $a$:    the time interval of data,  $b$: the number  of researchers, $c$: the number of publications,  $d$: the  number of journals,  $e$: the average number of researchers' publications, $f$: the average number of publications' authors. \end{flushleft}
\label{tab1}
\end{table}

\section*{The used features and mechanisms}

  \subsection*{Lotka's law and aging}

Lotka  analyzed the papers of   physics journals during the nineteenth century, and found  the  law: the number of papers    of a researcher approximately satisfies  that the number producing $n$ (where $n\in \mathrm{Z}^+$) papers is about $1/n^2$ of those producing
one\cite{Lotka1926}.   Price found    the inverse square law that   half of the publications come from the square root of all researchers\cite{Price1963}.
Lotka's law is
defined in the generalized form  $p(x=h)\propto h^a$,
where $a<-1$,  $h\in \mathbb{Z}$,   $x$ is random variable,
and
$p(x=h)$ represents the probability  that a researcher  published  $h$  papers.

With the Lotka's law, we can conclude  that  the probability  of a researcher  with $s$ publications at time interval $[T_0,T_1]$ in  a given dataset is proportional to $s^a$.
Assume that the number of publications of the researcher  at   $(T_1,T_2]$    is $cs^b$, where $b,c>0$.
 It gives rise to  $p(x=cs^b)\propto s^a$ at  $(T_1,T_2]$. Letting $h=cs^b$ obtains  $p(x=h)\propto h^{a/b}$.
 Therefore,   with this assumption,
   Lotka's law can hold at the following time interval. It  gives    reasonability  to assume
   that   the probability of a researcher, $i$, publishing papers at   $ t$ satisfies
\begin{equation}p_i(t)\propto  { (h_i(t-1) +1)^{\alpha }},\label{Lotka}\end{equation}
   where  $\alpha $ tunes the  effect of  cumulative advantage.
Larger values of $\alpha$ indicate higher probability   for productive researchers    to publish papers.

   Aging is  empirically observed
in   productivity patterns. Lehman concluded that productivity usually begins in a researcher's 20s, rises sharply to a peak in
the late 30s or early 40s, and then declines slowly\cite{Lehman2017}.
 The cumulative advantage and aging describe a  curvilinear function
for a researcher's publication productivity, rapidly increasing   and then slowly decreasing\cite{Simonton1984}.
Therefore, the right side of Eq.~(\ref{Lotka}) should be modified as
   \begin{equation}p_i(t)\propto (  h_i(t-1) +1)^{\alpha \mathrm{e}^{- \beta h_i(t-1)   }} \label{Lotka2},\end{equation}
     where  $\beta>0$ tunes the  effect of  aging. Larger values of $\beta$ indicate more quickly   researchers age.



\subsection*{Relation  between   publications and    coauthors}

\begin{figure*}[h]
\centering
\includegraphics[height=3.5   in,width=4.6     in,angle=0]{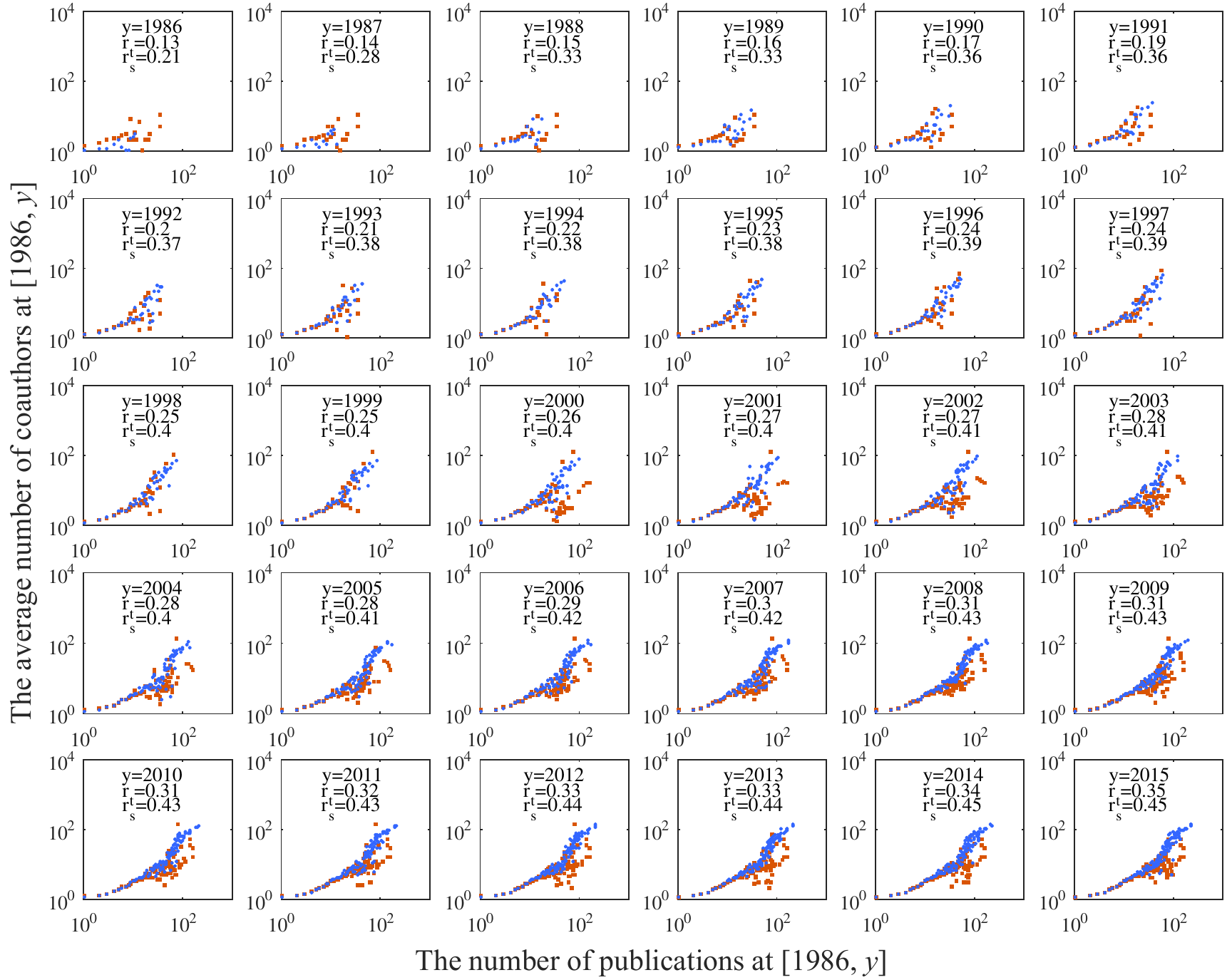}
\caption{   {\bf The correlation
between the number of  papers   and  the number of coauthors. }  Consider the
authors who  published papers  at $[1951,y]$, where $y=1992,...,2005$.
The panels show   the average number of coauthors
  at   $[1951,y]$ of  the authors with  the same  number of papers  at   $[1951,y]$ (red circles) and the predicted number (blue squares).
  The Spearman correlation coefficient, $r_t$,  for the empirical dataset  and that, $r_s$, for the synthetic dataset   are significantly larger than $0$, $p$-value~$<0.05$.}
  \label{deg_hd}
\end{figure*}

The positive correlation between the number of publications of a researcher and the number of his or her coauthors
has been found  in several empirical datasets\cite{Xie2019Scientometrics}. Fig.~\ref{deg_hd} shows the correlation  also appeared in the dblp dataset.   Arguments
 exist on   whether scientific  collaboration has a positive effect on publishing productivity. Lee et al found that the number of coauthors is not a significant predictor of the number of  publications\cite{Lee2005}. However,
  Ductor showed that after controlling for endogenous coauthorship formation, unobservable heterogeneity, and time, the effect of intellectual collaboration on the number of an individual's publications becomes positive\cite{DuctorL2015}.
 Fig.~\ref{deg_hd} also shows
that this correlation   in the dblp dataset  is not strong.
 Therefore, a variable   $u_i(t)$  is introduced to describe the potentiality of attracting researchers to coauthor\begin{equation}
u_i(t)= h_i(t)^\gamma   k_i(t)^{  1-\gamma }  ,
\label{Lotka3}
\end{equation}
where $ k_i(t)$ is the historical number of coauthors at $t$, and  $\gamma\in (0,1]$  tunes the inclination
for researchers to collaborate with
productive researchers or the researchers with many coauthors   in the past.

\subsection*{Cooperative game}

 A  cooperative game  consists of a set of players   and  a
characteristic function that specifies  a value to any  subset  of
players, e.g.,   the maximum of  benefit-cost ratio of that subset.  Coauthorship  can be regarded as a result  of that   game,
maximizing benefit and minimizing cost.
Researchers    can be regarded as players.    Coauthoring a researcher with a high reputation  contributes to  academic success\cite{QiMZengA2017}, and thus
 the reputation can be viewed as a kind of   benefit.
The  investments of  manpower and material resources
 on   research  can be regraded as  the  cost  of cooperation.

When modelling coauthorship networks by   geometric hypergraphs,
 and the distance
$d(i,j)$ between node $i$ and $j$   can be used to simulate   the   cost of researcher $j$ coauthoring with researcher $i$.
The value of $(u_i(t) +1)^{\mathrm{e}^{-\beta  u_i(t)  }}$
can be used to simulate   the benefit of  coauthoring with   $i$.
Then, the benefit-cost ratio of $j$ coauthoring with   $i$  can be modelled by
 \begin{equation}v_{i}(j,t)=\frac{1}{d(i,j)} (    u_i(t-1) +1)^{\alpha \mathrm{e}^{- \beta u_i(t-1)  }}\label{game}.\end{equation}

Consider the situation  that $j$ wants to write a paper as the principal author, which could be the first or the corresponding author. If we need to  assemble  a hyperedge $e$  for $j$, we  will sort  $v_{i}(j,t)$ for all $i$ from small to large, and choose the first $|e|-1$ nodes to coauthor with $j$.
The summation of the benefit-cost ratio of members in $e$ is the largest in the perspective of $j$, and this can be regraded as the  value of characteristic function on $e$.
This assembly mechanism will be used here to   simulate   the process of
 a researcher    finding coauthors.

  \subsection*{Invisible college and  isolated schools}

The analysis of team size does not address how
  teams  embed in a coauthorship
network.
 The embedding
way  in part reflects   the manner in
which researchers access the scientific community and the knowledge of their fields\cite{Burt2004}.
  Coauthorship networks usually have
  giant components comprising many nodes, as well as many small components.
Giant components would be
supporting evidence for the  invisible
college,  a  community of  researchers who often   exchange  ideas and encouraged each other.
Small components would be
supporting evidence for isolated schools. That is, many   teams   are
 likely to draw from
different scientific communities.


Lotka's law and maximizing  benefit-cost ratio cannot generate  giant components in a network, which needs the collaborations among  researches from different universities and countries.
 Researchers   found some   possible factors of these collaborations, such as the institutional prestige\cite{Hunter-Leahey2008},   and academic  performances   of researchers\cite{Aldieri-Kotsemir2018}.
Therefore, the reputation of a researcher      given by  his or her institutional prestige or academic  performance
is a possible  factor that drives the formulation of giant components.
Therefore, we will introduce a variable called reputation to the model.

\begin{figure*}[h]
\centering
\includegraphics[height=3.5   in,width=4.6     in,angle=0]{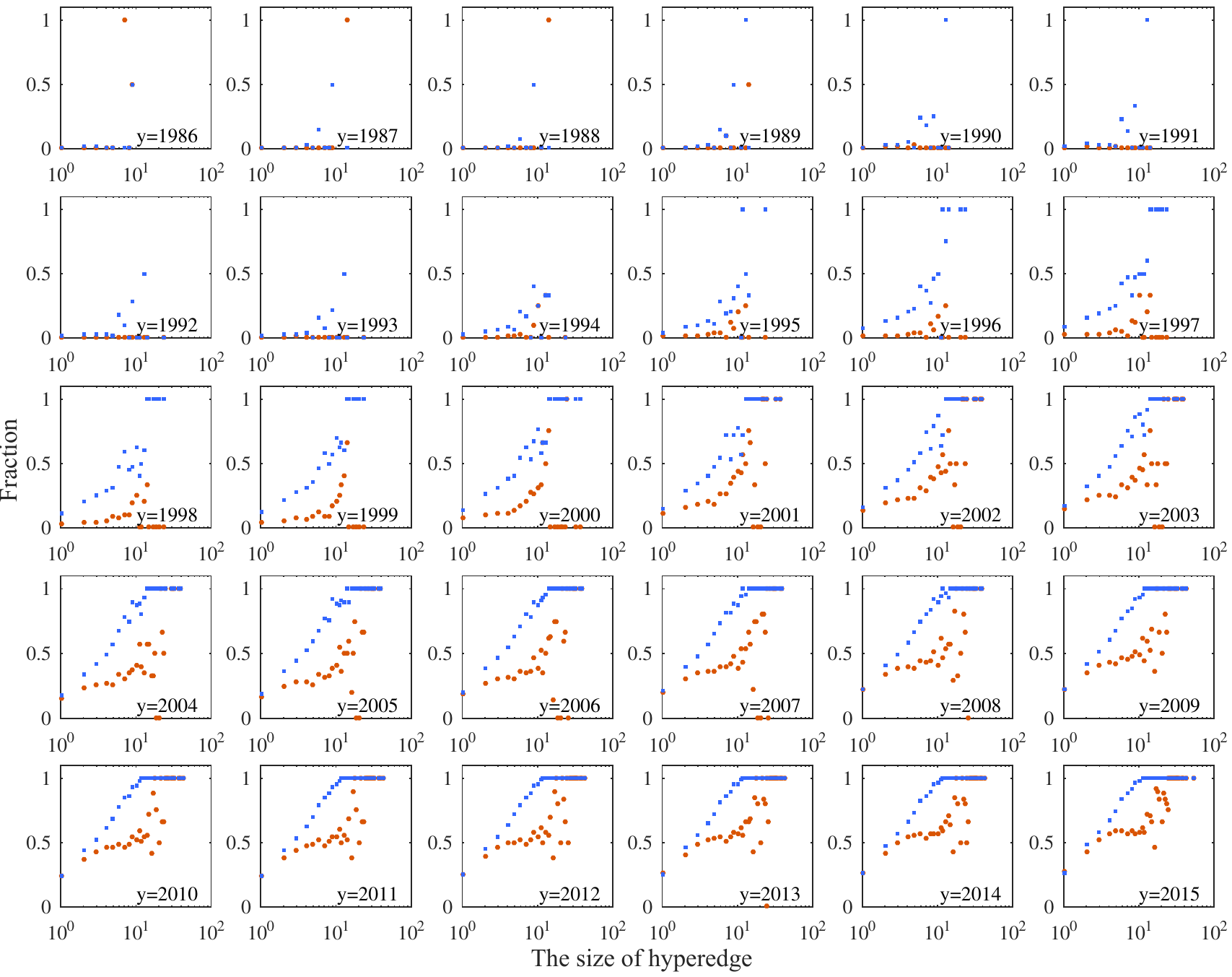}
 \caption{        {\bf  The proportion of hyperedges belonging to a giant component.}
 The red dots show the proportion for hyperedges with a given size of the empirical dataset at $[1986, y]$ (red circles) and the predicted proportion (blue squares).           }
    \label{prob_giant}
\end{figure*}

  The model will replace  the last members of some  hyperedges by   other nodes with  high reputations.
Fig.~\ref{prob_giant}
 shows    the proportion of hyperedges with a given size   belong to a giant component.
It indicates that the proportion increases with the growth of hyperedge size    and time. Therefore, in the model, the
    probability of selecting a hyperedge to replace its last member  is designed to increase with   its  size   and time. The replacing mechanism guarantees the emergence of   giant components in synthetic datasets.


  \section*{The model}

  \subsection*{Distributed design}
Some researchers and teams publish papers concurrently, with no effect  to each other. That is, the system of researchers and their collaborations evolves  in a parallel mode; thus it   should be simulated by a     distributed system that locates its components   on different networked computers. The components communicate and coordinate their actions by passing messages to others, interacting with  others in order to achieve a common goal\cite{Kshemkalyani2008}. Three significant characteristics of distributed systems are: concurrency of components, lack of a global clock, and independent failure of components.
The system of researchers collaborating also has  these characteristics.



In a distributed system, a message has three   parts: the sender, the recipient, and the content. The sender needs to be specified so that the recipient knows which components sent the message, and where to send replies.
 For the distributed model of collaborations here, the message includes  the reputation  of researchers, their number of papers, and their number of coauthors   changed   in each process.
Each  process acts  as both the sender  and     recipient.
   The impacts of the changes of these variables    need time to propagate from local to global, and   may  have a lag.
Therefore, the message passing does not   need to be timely.
 In the simulation here,  the processes send the message  to others
  yearly (the unit of the time in the model is year).
 At each time step,    the changes made by a process only affect the computation in itself.
 The computing  cost of passing message among  processes is reduced.

\subsection*{Mathematical formulae}

The number of publications of a researcher is easily affected by random factors
from his or her work environment and family.
Therefore,  we
 draw   $x_s(t-1)$ from $\mathrm{Pois} \left((u_i(t-1) +1)^{\alpha \mathrm{e}^{- \beta u_i(t-1)   }}\right)$ for any existing  node $s$.
 In the model, for each process and each new hyperedge at time $t$, we will draw a node $i$ to assemble that  hyperedge as the
principal member according to a     probability
    \begin{equation}p_i(t )= \frac{x_i(t-1)}{\sum_sx_s(t-1)}.\label{Lotka4}\end{equation}
Since the degree and hyperdegree are positive correlated, we only analyzed the case $\gamma=1$ in following discussion for simplicity.

\begin{figure*}[h]
\centering
\includegraphics[height=1.5    in,width=4.5     in,angle=0]{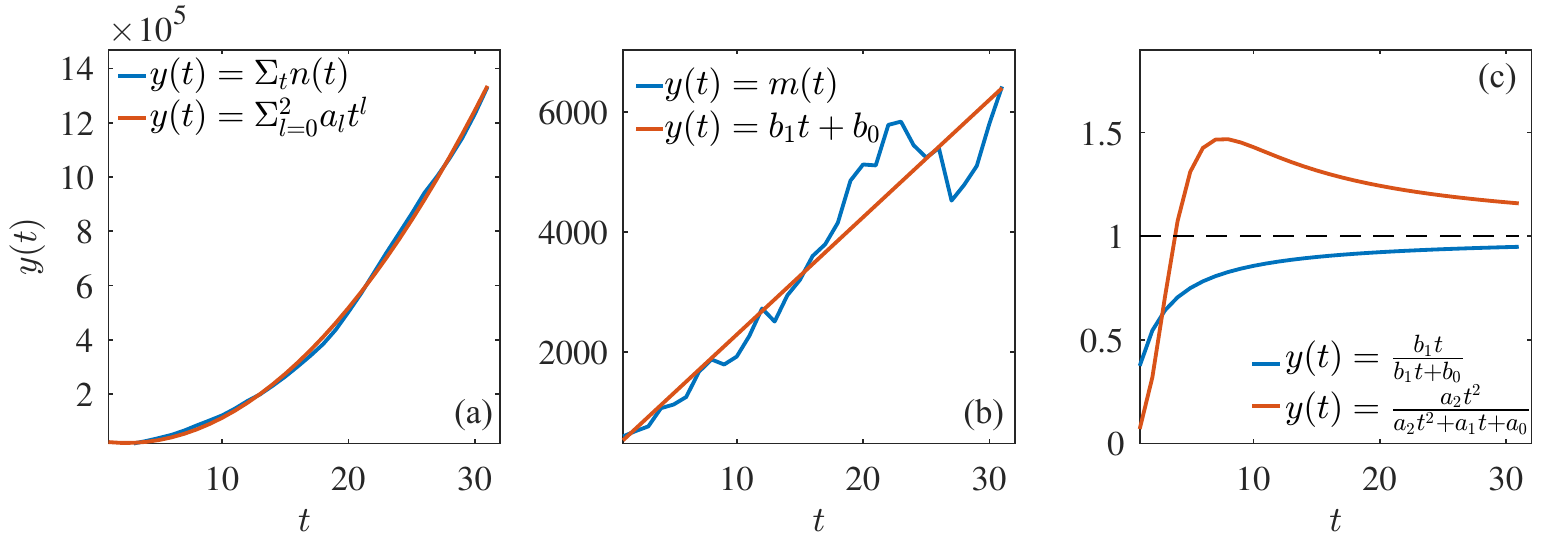}
\caption{   {\bf  The increasing trend  of data size.} Panel  (a) shows the trend for nodes and its  fitting polynomial  $y(t)=\sum^2_{l=0}a_lt^l$, where  $a_0=2.931$e+04,  $a_1=-7.770$e+03,  and  $a_2=1.610$e+03.
Panel (b) shows that the trend for hyperedges and its  fitting polynomial $y(t)=\sum^1_{l=0}b_lt^l$, where
   $b_1=327.6$ and  $b_2=196.1$. Panel (c) shows the contributions of the leading terms to these polynomials.
   }
  \label{number_trend}
\end{figure*}

Firstly, we consider
the situation that an author $i$   writes  a paper as the principal  author. Fig.~\ref{number_trend} shows     the fitting polynomials of the number of  nodes and that of hyperedges, which are dominated by their leading terms.
Therefore, when   $\beta(h_i(t-1)+1)\ll1$,  the increment of hyperdegree  of $i$    as the principal member
  \begin{equation} \Delta_1 h_i(t ) \approx b_1t p_i(t ) \approx \frac{b_1t (h_i(t-1) +1)^{\alpha } }{   a_1 t^2 ( \sum_{h } q_hh^\alpha)}  = \frac{  \lambda_1 }{t} (h_i(t-1) +1)^{\alpha } ,\label{Lotka5}\end{equation}
  where $\lambda_1={b_1  }/{\left(   a_1  \sum_{h } q_hh^\alpha  \right)} $. The value of  $\sum_{h } q_hh^\alpha $ is a finite constant due to the proportion of $h$-hyperdegree nodes $q_h\propto h^{-3}$ for the synthetic dataset, which will be shown in following sections.

\begin{figure*}[h]
\centering
\includegraphics[height=1.5    in,width=4.5     in,angle=0]{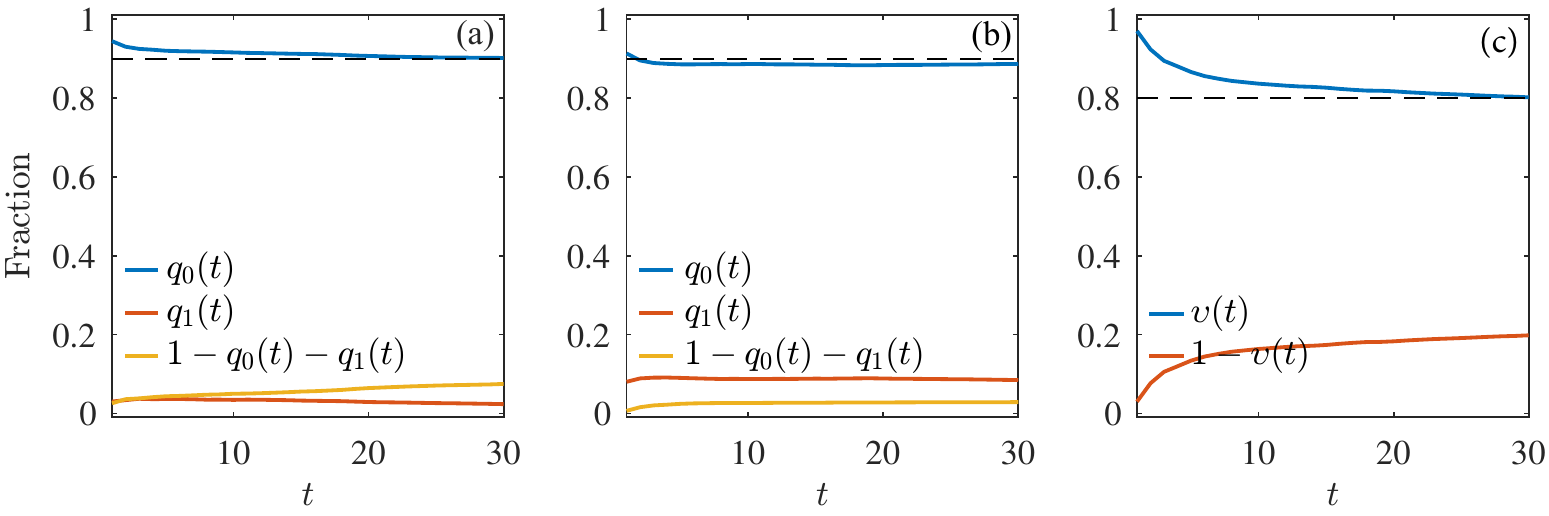}
\caption{   {\bf  The compositions of  nodes in the synthetic dataset.} The panel  (a) shows
the proportion, $q_0(t)$,  of $0$-degree  nodes at each time step $t$,  and the proportion, $q_1(t)$,   of $1$-degree  nodes at each time step $t$.
The panel (b) shows these proportions for hyperdegree.
The panel (c) shows   the proportion, $\upsilon(t)$, contributed by nodes with hyperdegree no larger than $1$ in   $\sum_ih_i(t)^\alpha$, where $\alpha=1.43$.}
  \label{rate_hyper}
\end{figure*}

Secondly, we consider
the situation that   an author $i$  writes a paper as a team member.
In the model, we consider a node $j$ with distance $d(i,j)$ from $i$.
Then the expected number of nodes between them is approximately equal to  $d(i,j)  a_2 t^2/(2\pi) $.
The expected distance of the closest node to $j$ is approximately equal to $2\pi/(a_2 t^2)$. Fig.~\ref{rate_hyper} shows
in  the synthetic  dataset,  there are more than $q_0=90\%$ nodes with $0$-hyperdegree   at each time step. Therefore, we  can approximately regard
  all of the nodes drawn between $i$ and $j$ as    $0$-hyperdegree nodes.
Then, if node $j$ is chosen to assemble a hyperedge by Eq.~(\ref{Lotka4})  as the principal member,  node $j$ will choose a node  $i$ with    $d(i,j)< 2\pi(    h_i(t-1) +1)^{\alpha \mathrm{e}^{- \beta h_i(t-1)  }}/(a_2 t^2) $.
 Table~\ref{tab1} shows that  the average size of hyperedges  is less than $3$; thus we can  assume  that $j$ only chooses  one node.
 Therefore, the increment of hyperdegree   of $i$    as a team member  is
 \begin{equation}\Delta_2h_i(t) \approx   \frac{2\pi  b_1t} { a_2 t^2 } (    h_i(t-1) +1)^{\alpha \mathrm{e}^{- \beta h_i(t-1)  }}\approx  \frac{\lambda_2} {t} (    h_i(t-1) +1)^{\alpha \mathrm{e}^{- \beta h_i(t-1)  }},
 \label{game_c}\end{equation}
 where $\lambda_2=2\pi b_1 /a_2$.

%

Eq.~(\ref{Lotka5}) and Eq.~(\ref{game_c}) give  rise to
\begin{equation}\frac{d}{dt}h_i(t) =    \frac{\lambda }{  t}  (h_i(t)+1)^{\alpha  \mathrm{e}^{- \beta h_i(t)  }}
\approx  \frac{\lambda }{  t}  (h_i(t)+1)^{\alpha  }
\approx     \frac{\alpha\lambda}{  t}  \left(h_i(t)+\frac{1}{\alpha}\right)    , \label{Lotka6}
\end{equation}
where $\lambda=\lambda_1+\lambda_2$, and $(h_i(t ) +1)^{\alpha }$ is approximated by its first order  of Taylor series.
Note that this approximation cannot be suitable for a large $t$, and $t\leq 30$ here. Let $t_i$ be the time when $i$   generate.
 The solution to Eq.~(\ref{Lotka6})
gives   node $i$'s expected hyperdegree   $\bar{h}_i(T)=(T/t_i)^{\alpha\lambda}-1/\alpha$, which   yields   $p(\bar{h}_i(T)\leq h) =  p (t_i \geq    T/(h+1/\alpha)^{1/(\alpha\lambda)}  ) $.
Fig.~\ref{number_trend} shows the cumulative number of nodes can be fitted by   $y(t)=\sum^2_{l=0}a_lt^l$; thus
 $p (t_i<t)=(t-1)(t-2)/T(T-1)\approx t^2/T^2$. Hence
 $ p (t_i   \geq    T (h+1/\alpha)^{-1/(\alpha\lambda)} ) =1-p (t_i  <    T (h+1/\alpha)^ {-1 /( \alpha\lambda}) )\approx 1-   (h+1/\alpha)^{- {2}/{(\alpha\lambda)}} $. It gives rise to
 \begin{equation} p(h)=\frac{d }{d h}p(\bar{h}_i(T) \leq   h) \propto  \left(h+\frac{1}{\alpha}\right)^{-1-\frac{2}{\alpha\lambda}}\approx h^{-1-\frac{2}{\alpha\lambda}},\label{Lotka7}
\end{equation}
which shows
the reason for the emergence of   the   power-law part of hyperdegree distribution.


When $\beta(h_i(t)+1)\gg1$,  Eq.~(\ref{Lotka6}) gives rise to
\begin{equation}\frac{d}{dt}h_i(t) \approx  \frac{\lambda}{  t}  . \label{exp}
\end{equation}
The solution to Eq.~(\ref{exp})
gives rise to  $\bar{h}_i(T)= \lambda \log (T/t^*_i)+C_i  $,
where $C_i$ is the    hyperdegree accumulated from  the process that Eq.~(\ref{exp}) does not hold,
and $t^*_i$ is the start time that Eq.~(\ref{exp}) holds.
It   yields   $p(\bar{h}_i(T)\leq h) =
  p (t^*_i \geq    T \mathrm{e}^{-  (h-C_i)/\lambda}  ) $.
Since $C_i$ satisfies $\beta(C_i+1)\gg1$ for any possible $i$; thus there exists a constant $C$ as the minimum of those $C_i$, such that  $h_i(t)$ is controlled by Eq.~(\ref{exp}) when    $h_i(t)\geq C$.
 The value of $C$ is mainly  accumulated from  the process  controlled by Eq.~(\ref{Lotka6});
thus
$C\approx (t^*_i/t_i)^{\alpha\lambda}-1/\alpha$, and then   $t^*_i\approx(C+1/\alpha)^{1/(\alpha\lambda)}t_i$.
 Hence
 $p (t^*_i  \geq T \mathrm{e}^{- (h-C)/ \lambda }  ) =1-p (t^*_i  <    T \mathrm{e}^{-  (h-C)/\lambda} )\approx1- (C+1/\alpha)^{-2/(\alpha\lambda)} \mathrm{e}^{-  2(h-C)/\lambda}
$. It gives rise to \begin{equation}p(h)\approx\frac{d}{dh} p(\bar{h}_i(T) \leq   h) \propto \frac{2}{\lambda}  \mathrm{e}^{- \frac{2}{\lambda}(h-C)} ,\label{exp2}
\end{equation} which  is an exponential distribution on the interval $[C,+\infty)$.
 Therefore, we can expect  a   power-law distribution with an exponential cutoff  for     hyperdegrees.

\begin{figure*}[h]
\centering
\includegraphics[height=3.5    in,width=4.6     in,angle=0]{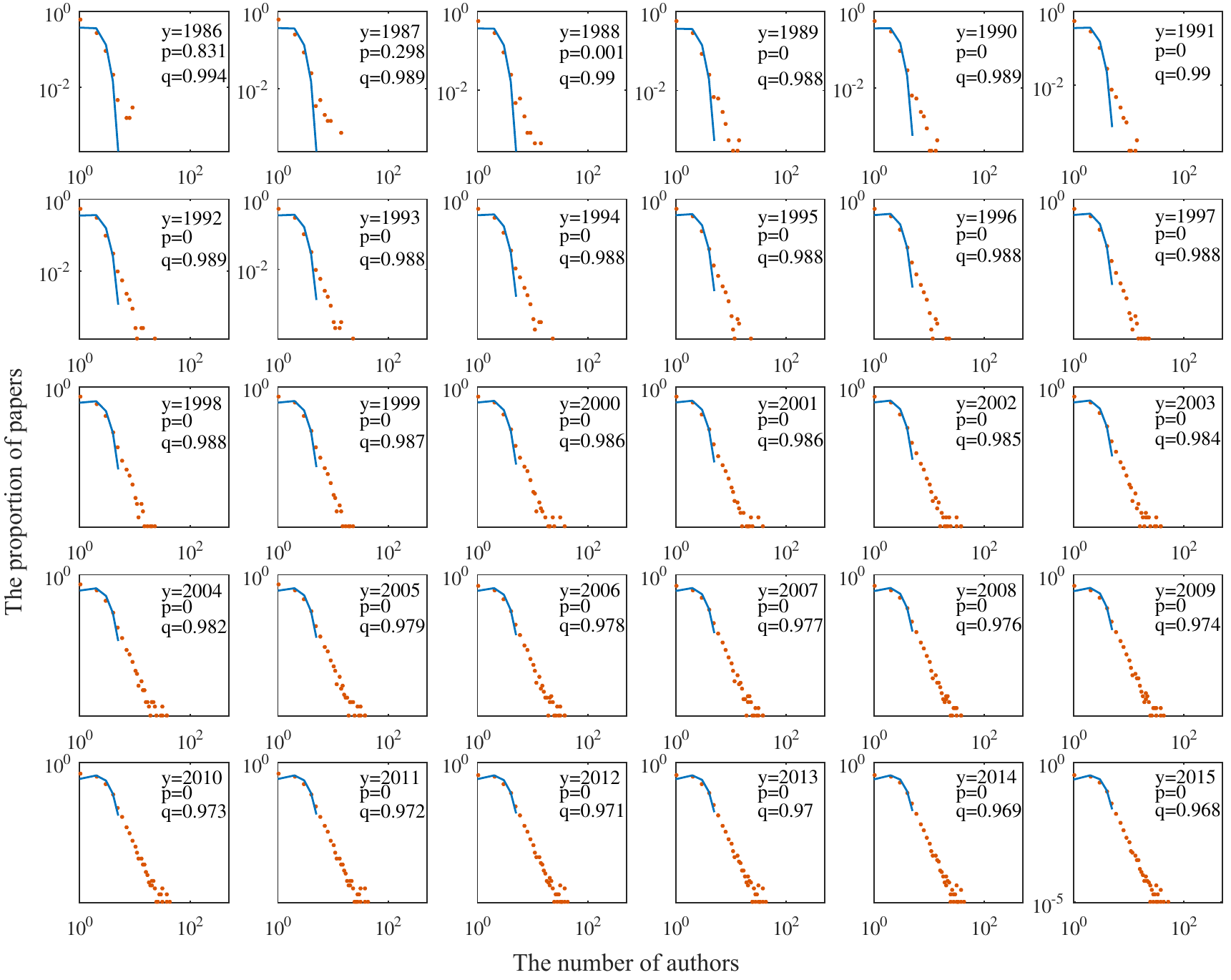}
\caption{ {\bf The distribution of the size of hyperedge.}  The panels show that the  distribution  (red circles) has a  hook head, close to a   Poisson distribution (blue curves).
Index $p$ is the $p$-value of KS test for the null hypothesis that the size distribution for
   the hyperedges with sizes smaller than $7$  is a Poisson distribution.
When $p<0.05$, the test rejects the null hypothesis at the 5\% significance level, and cannot otherwise.
Index $q$
is the proportion of those hyperedges.  }
  \label{hyperedge}
\end{figure*}

\begin{figure}\centering
\includegraphics[height=3.     in,width=3.3    in,angle=0]{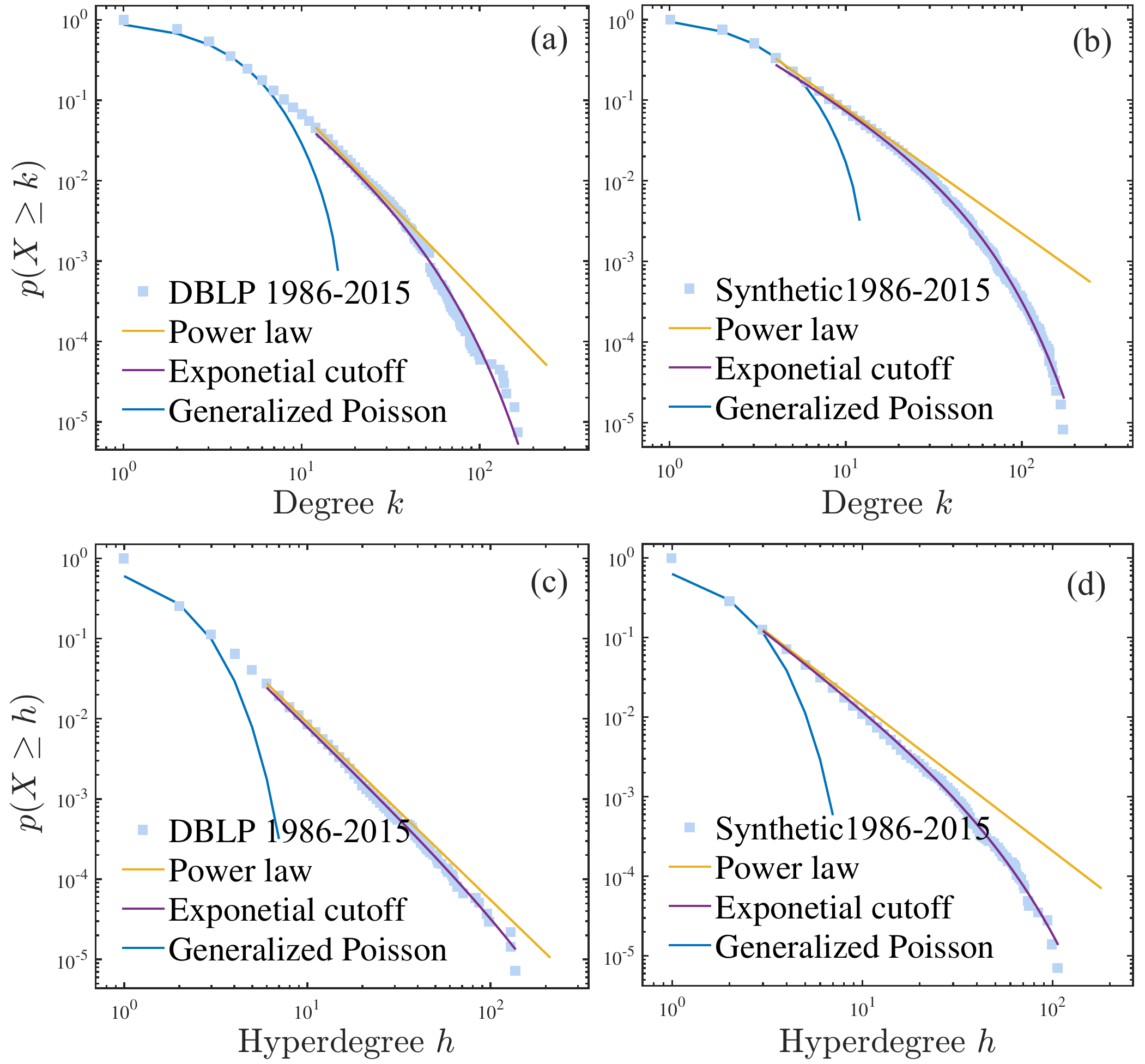}
  \caption{ {\bf The  hyperdegree  and degree distributions.} The panels  show
  the trichotomy of degree and hyperdegree
distributions:
 a generalized Poisson head, a power-law   midsection, and an exponential
cutoff.  }
 \label{tri_degree}      
\end{figure}

 Due to the positive correlation between degree and hyperdegree, we can also   expect  a   power-law distribution with an exponential cutoff  for      degrees.
Now we turn to the hook head of degree distribution.
 Milojevi\'c studied the empirical datasets from the disciplines of astronomy, literature
and social psychology. She found that the  distribution of the number of authors of a paper
is  well fitted by a mixture of  two Poisson distributions  and a truncated power law\cite{Milojevic2014}.
For the   dblp dataset,
there are 70.0\% nodes with  hyperdegree     one.
 Fig.~\ref{hyperedge}   shows that the head of
the distribution of the size of hyperedge  has a close shape to a    Poisson distribution.
Those generate the hook head  of degree distribution.
Fig.~\ref{tri_degree} shows the  analyzed   features   of the degree and hyperdegree distributions for the empirical and synthetic datasets.



  \subsection*{Implementation }

\begin{figure*}[h]
\centering
\includegraphics[height=1.7    in,width=4.6    in,angle=0]{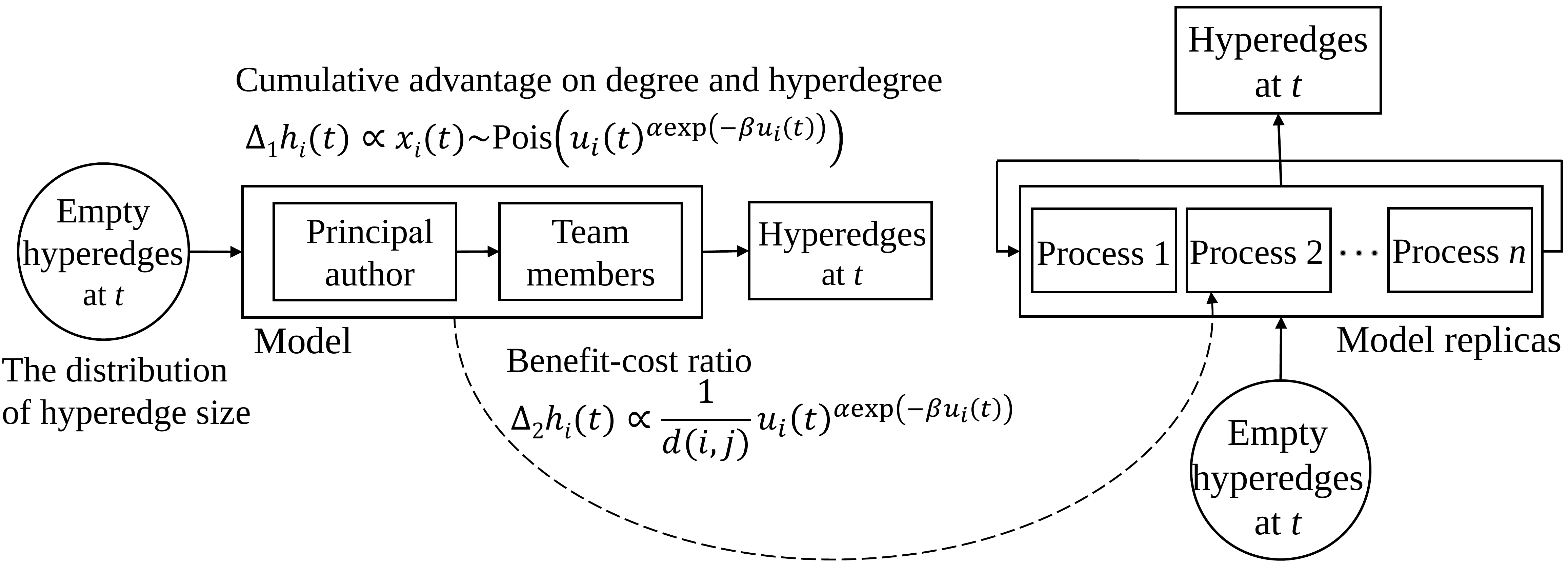}
\caption{   {\bf   An illustration of the distributed hypergraph  model.}
Authors are represented by nodes.
The set of authors of a paper is represented  by a hyperedge. The number of hyperedges and their size, the inputs  of the model, are the same to those the empirical data    (round frame).   }
  \label{illus}
\end{figure*}

We build  a model      on  a circle $S^1$, and express the cost $d(i,j)$  in Eq.~(\ref{game}) by the arc-length between node $i$ and $j$.
We run the model from $t=1$ to $30$, only 30   steps, simulating the evolution of the dblp dataset from 1986 to 2015. That is, the time in the model is that in reality.
The model generates
new hyperedges at each time step.  Fig.~\ref{illus} shows the illustration of the proposed model. 

We start at time zero with $n(0)$ nodes that  are sprinkled
on $S^1$   uniformly and randomly,
where $n(0)$  is the number of  the authors who have papers at time interval $[T_0,T_1]$.
 When a node  is sprinkled, its spatial position is fixed.
Let $h_i(0)$ and $k_i(0)$ be the  historical hyperdegree and degree of an  author $i$ in the empirical  dataset  at  $[T_0,T_1]$.

At each time step $t$,  we sprinkle  $\varepsilon    n(t)$
new nodes  on $S^1$   uniformly and randomly, where $\varepsilon>1$, and  $n(t)$ is the number of new authors who appear at $(T_{t-1},T_t]$.
Then, we assemble $m(l)$ new hyperedges, where $m(l)$  is the number of the
papers  at $(T_{t-1},T_t]$. The sizes of hyperedges are  the same to those of the empirical   data.
For each hyperedge,
 we select  a node   as its principal member according to  the probability in Eq.~(\ref{Lotka4}),
 and then we select the rest members
 according to the decreasing rank given by  Eq.~(\ref{game}).

  The assembly mechanism
 simulates the process of a researcher  seeking for the collaborators with enough experiments  of
publishing and collaborating (modelled by $u_i(t)$) and for   a small cost for cooperation (modelled by $d(i,j)$).
Researchers and  teams can publish   papers     simultaneously with no or small effect on each other. Therefore, we     assemble nodes as hyperedges by   distributed computing.
 Algorithm~\ref{alg1} shows how to run the model in a distributed way.


\begin{algorithm}
\caption{The distributed model }
\label{alg1}
\begin{algorithmic}
\REQUIRE ~~ 
\\the annual number of new nodes   $\varepsilon n(t)$, $t=1,...,T$;
\\the  empty hyperedges with given   sizes;
\\the historical degree    and   hyperdegree of the nodes at $t=0$.
\ENSURE ~~a hypergraph.\\ 
\STATE{ Sprinkle   $n(0)$ nodes    on $S^1$ uniformly and randomly;}
\STATE{ initialize reputation $r_i(0)=u_i(0)$;}
  \FOR{$t$  from $1$ to  $T$  }
 \STATE{sprinkle $\varepsilon n(t)$ new nodes   on $S^1$ uniformly and randomly and initial their reputation $0$;}
 \STATE{subdivide  the empty  hyperedges at time $t$  into  subsets $\{S_w| w=1,..., W\}$;}
 \FOR{  job $w$    from $1$ to  $W$ }
 \FOR{ each empty hyperedge $e$ in   $S_w$ }
 \STATE{choose  the principal member  according to the probability  in Eq.~(\ref{Lotka4});}
 \STATE{choose the first $|e_l|-1$ nodes  according to the decreasing  rank given by Eq.~(\ref{game});}
 \STATE{sample  $x$ from    $U(0,1)$; }
\IF{$x<\epsilon|e|t$ and $|e|>1$}
\STATE{replace the last member   by Algorithm~\ref{alg2};}
\ENDIF
\ENDFOR\ENDFOR
\STATE{add hyperedges to the hypergraph;}
\STATE{update   degree,   hyperdegree,     and   reputation.}
\ENDFOR
\end{algorithmic}
 \end{algorithm}



The distance $d(i,j)$ makes the nodes mainly connect the nodes   nearby.
 This will generate  a large fraction of small components. Algorithm~\ref{alg2} is proposed to   generate  giant components by   replacing the  last member of some hyperedges by another node.
It makes   nodes  connect to the neighbors of  the nodes with high reputations. This design  aligns
with the common sense   that
leaders of some famous research teams may receive many collaboration invitations, and they
 would arrange their team members to follow up.
 The design of the   probability,  $\epsilon|e|l$, of exchanging members is based on the evidence of empirical dataset  that   the more members the higher probability the team connecting to the giant component, and the probability increases with time.


\begin{algorithm}
\caption{Generating giant components}
\label{alg2}
\begin{algorithmic}
\REQUIRE ~~ $k_i(t-1)$, $h_i(t-1)$, and $r_i(t-1)$;
\IF{$x<\epsilon   |e| t$: }
\STATE{let $y_s(t-1)=k_s(t-1)+r_s(t-1)$ for any node $s$;}
\STATE{ select  a node, $i \notin e$ according to  the probability  $ { y_i(t-1)^{\alpha \mathrm{e}^{- \beta y_i(t-1) }}}/{\sum_s y_s(t-1)^{\alpha \mathrm{e}^{- \beta y_s(t-1)  }}}$ ;}
\STATE{update   $r_i(t)=r_i(t-1)+1$;  }
 \STATE{     select a node $j$ randomly from $i$ and its neighbors$~\notin e$; }
      \STATE{ replace the
  last member of $e$ by $j$.  }
   \ENDIF
\end{algorithmic}
 \end{algorithm}







\section*{Results}
\subsection*{Parameters and statistical   indexes}

The number of new  hyperedges at each time step is the annual  number of papers in Set 2, and the size of a hyperedge
is the number of authors of a paper.
The number of new  nodes at each time step is proportional to the    number of the    authors   appearing in  Set 2   at that time, where the proportion $\varepsilon=9$ here.
This is because some nodes generated by   the model will not join  any hyperedge. Those nodes  are used to express  the researchers who have  no papers right now  but  have the potential to publish. We used  $W=48$  processes to run the model.

We explored the parameter spaces of the model.
  When  $\alpha=1.43$,  $\beta=0.0016$, and $\gamma=0.9$,  the model  can generate hyperdegree   and degree distributions with  close shapes  to    the empirical  distributions.
 The   $\gamma$ takeing value $0.1$ indicates     the inclination of connecting nodes with a high hyperdegree not high degree
 Letting  $\epsilon=0.0036$  can  make
the difference between the size  of the giant component of
DBLP 1986-2015 and  the predicted size is smaller than 0.1.

\begin{figure*}[h]
\centering
\includegraphics[height=1.2    in,width=4.7     in,angle=0]{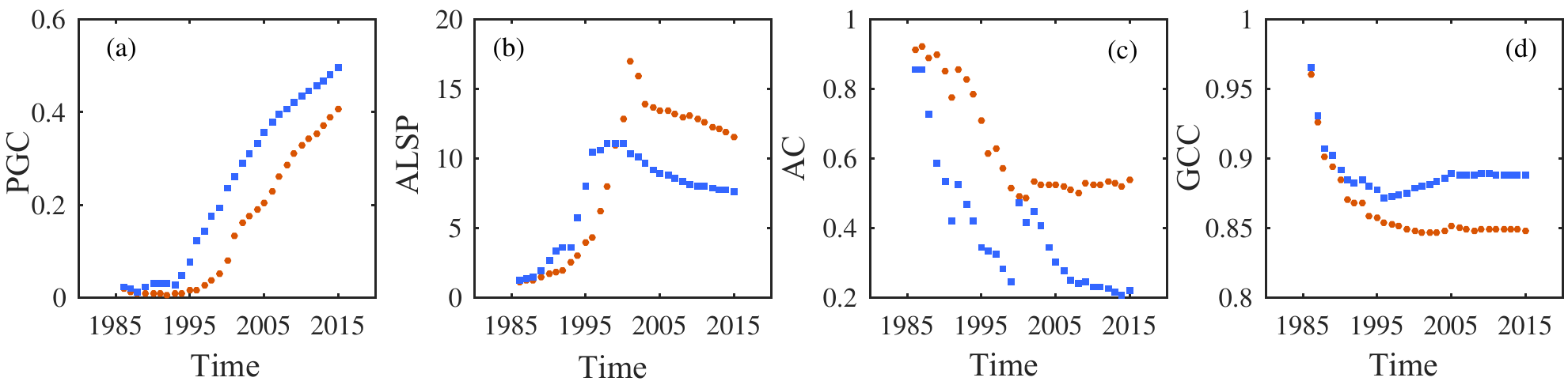}
\caption{    {\bf The evolutionary trends of four topological  features of the empirical and synthetic datasets.  }
The panels show the trend for the proportion of giant component (PGC), the average length of shortest paths (ALSP), the assortativity coefficient (AC), and global clustering coefficient (GCC). }
  \label{net_feat}
\end{figure*}

Table~\ref{tab1} shows certain statistical indexes of empirical  and synthetic datasets. Note that there are substantial differences  in   degree assortativity coefficient    and the average length of shortest paths.
To improve the agreement  in the first index, we can introduce a mechanism  to  connect  nodes with similar degrees, replacing the connects from nodes with small degrees to nodes with large degrees.
 It mitigates the role of nodes with large degrees that   make network connect; thus  may  increase  the average length of the shortest paths.
However, for the   simplicity of model, we do not add the  mechanism   here.
Fig.~\ref{net_feat} shows there is a big difference on the evolutionary trend of   assortativity coefficient.
 Whether it is caused
by    the model itself or    other reasons is
a question that we cannot answer conclusively.


 \begin{table*}[!ht] \centering \caption{{\bf Statistical   indexes of the  empirical and synthetic networks.} }
\begin{tabular}{l r r r r r r r r r} \hline
Network&NN&NE   & GCC & AC &ALSP  &  PGC  &MOD   \\ \hline
DBLP 1986-1995 &16,792& 16,049& 0.857   & 0.711 &  3.021   &0.012  &  0.803 \\
Synthetic 1986-1995 &16,626 &15,565  &0.877&0.346 &7.972 &0.056  &0.915\\
\hline
 DBLP 1986-2005  &64,287  & 91,132&   0.851 & 0.524   &13.422  &0.172 & 0.941 \\
Synthetic 1986-2005 &66,415  & 88,532 &0.889&0.301  &8.876 &0.278  &0.960  \\
\hline
  DBLP 1986-2015  &148,516&262,030 &0.848   &0.539 &11.556   &0.367&  0.956   \\
Synthetic 1986-2015 & 150,920 & 255,489 &0.889   & 0.220  &7.635 &  0.421  &0.940 \\
\hline
 \end{tabular}
  \begin{flushleft} The indexes are    the number  of nodes (NN), the number of  edges (NE),  global  clustering coefficient (GCC),  assortativity coefficient   (AC),  the average shortest path length (ALSP),      the   proportion of   giant component~(PGC), and modularity (MOD).
\end{flushleft}
\label{tab1}
\end{table*}

\subsection*{Collaboration patterns}



\begin{figure*}[h]
\centering
\includegraphics[height=3.5   in,width=4.6     in,angle=0]{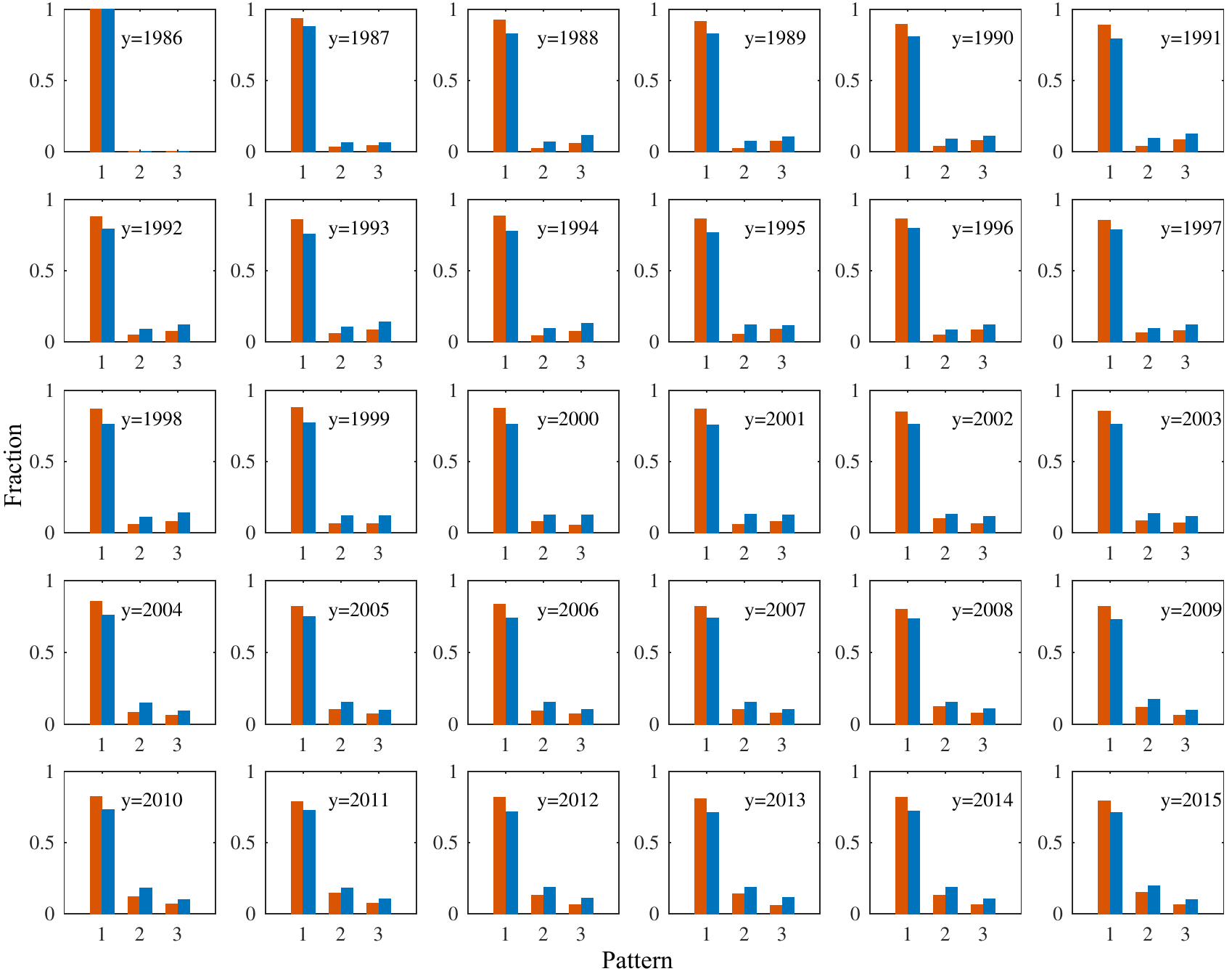}
\caption{    {\bf Collaboration patterns  from the perspective of multinary relationship.  }
Pattern 1 is that the authors of a paper never coauthored.
Pattern 2 is that  parts of those authors have coauthored, not all together.
Pattern 3 is that all of those authors have published  papers together.
The panels show the fractions of those patterns (red bars) and the predicted fractions (blue bars).
  }
  \label{pattern1}
\end{figure*}

From the perspective of multinary relationship, collaborations  can be
classified into three   patterns   according to the historical coauthorship of its authors: all new, partially new, and all old.
   The pattern  all new means that  its  authors never coauthored  before in  a given reference  dataset.
The pattern
partially new means that   parts of its authors,  not all,  have  coauthored before.  That is, some parts  of the multinary relationship
have appeared before.
  The pattern all old means that all of its authors have published papers together. That is, the multinary relationship has appeared before.
 Fig.~\ref{pattern1} shows that  the proposed  model  generate the three patterns. 

\begin{figure*}[h]
\centering
\includegraphics[height=3.5   in,width=4.6     in,angle=0]{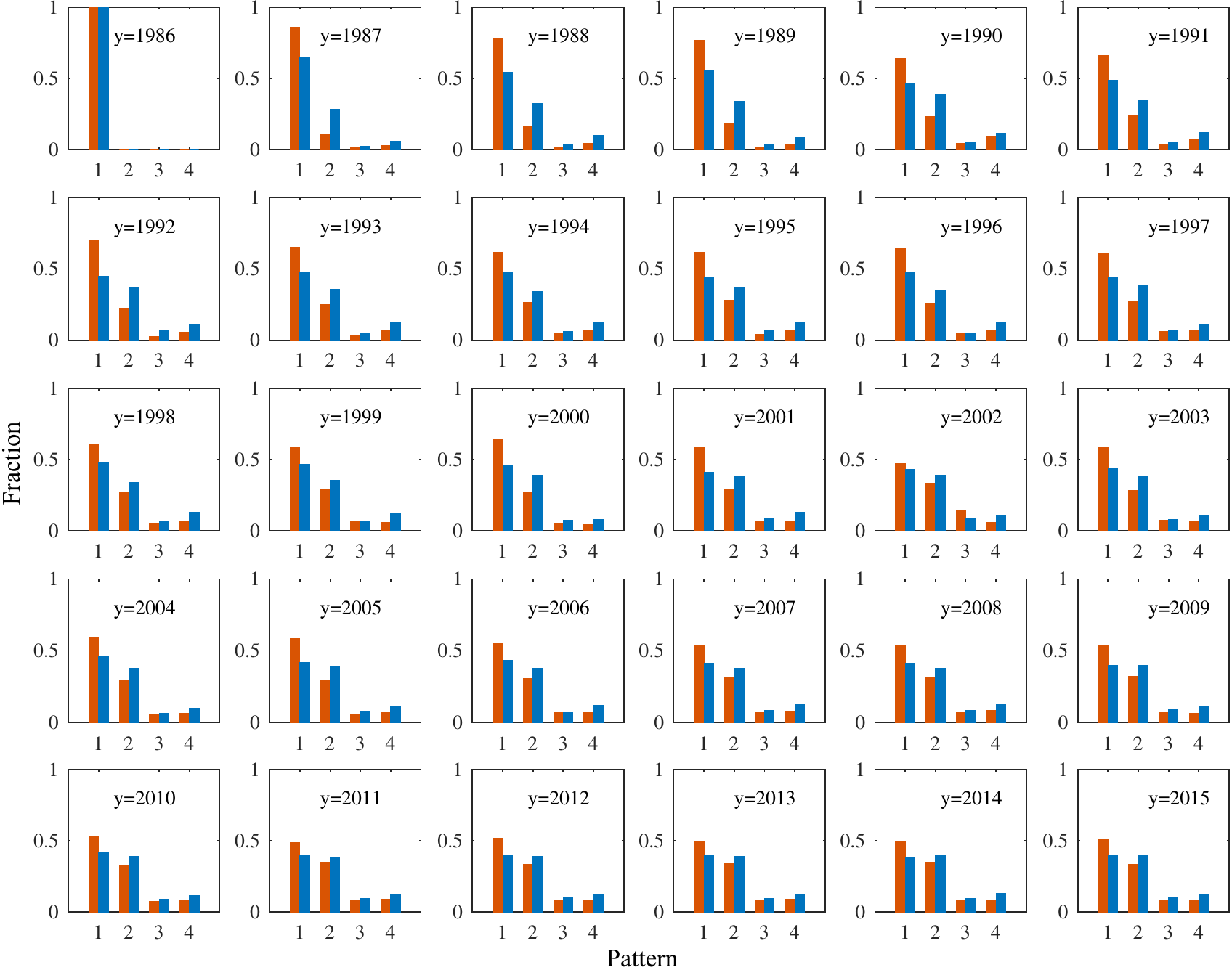}
\caption{    {\bf Collaboration patterns from the perspective of binary relationship.  }
Pattern 1 is that two authors, who have no paper in the dataset,  coauthored for the first time.
Pattern 2 is that one author have no paper in the dataset and the other have, and they coauthored for the first time.
Pattern 3 is that two author, who have  papers in the dataset,   coauthored for the first time.
Pattern 4 is that two author have  coauthored before in the dataset.
The panels show the fractions of  those patterns (red bars) and the predicted fractions (blue bars).
  }
  \label{pattern2}
\end{figure*}

From the perspective of binary relationship,
  collaborations    between two authors can be
  classified into four patterns according to the time of the authors appearing in  a given reference dataset:  new-new, new-old, old-old but have not coauthored yet, and old-old have coauthored before\cite{Guimera-Uzzi2005}.
  The pattern   new-new   means that   both authors     newly appear   in the dataset,
 and new-old   means that  one author   newly appears.  The meanings of the other two patterns are obvious.
 Fig.~\ref{pattern2} shows  that the four patterns can also be generated by the proposed  model.

\subsection*{Degree and hyperdegree distributions}

\begin{figure*}[h]
\centering
\includegraphics[height=3.5   in,width=4.6     in,angle=0]{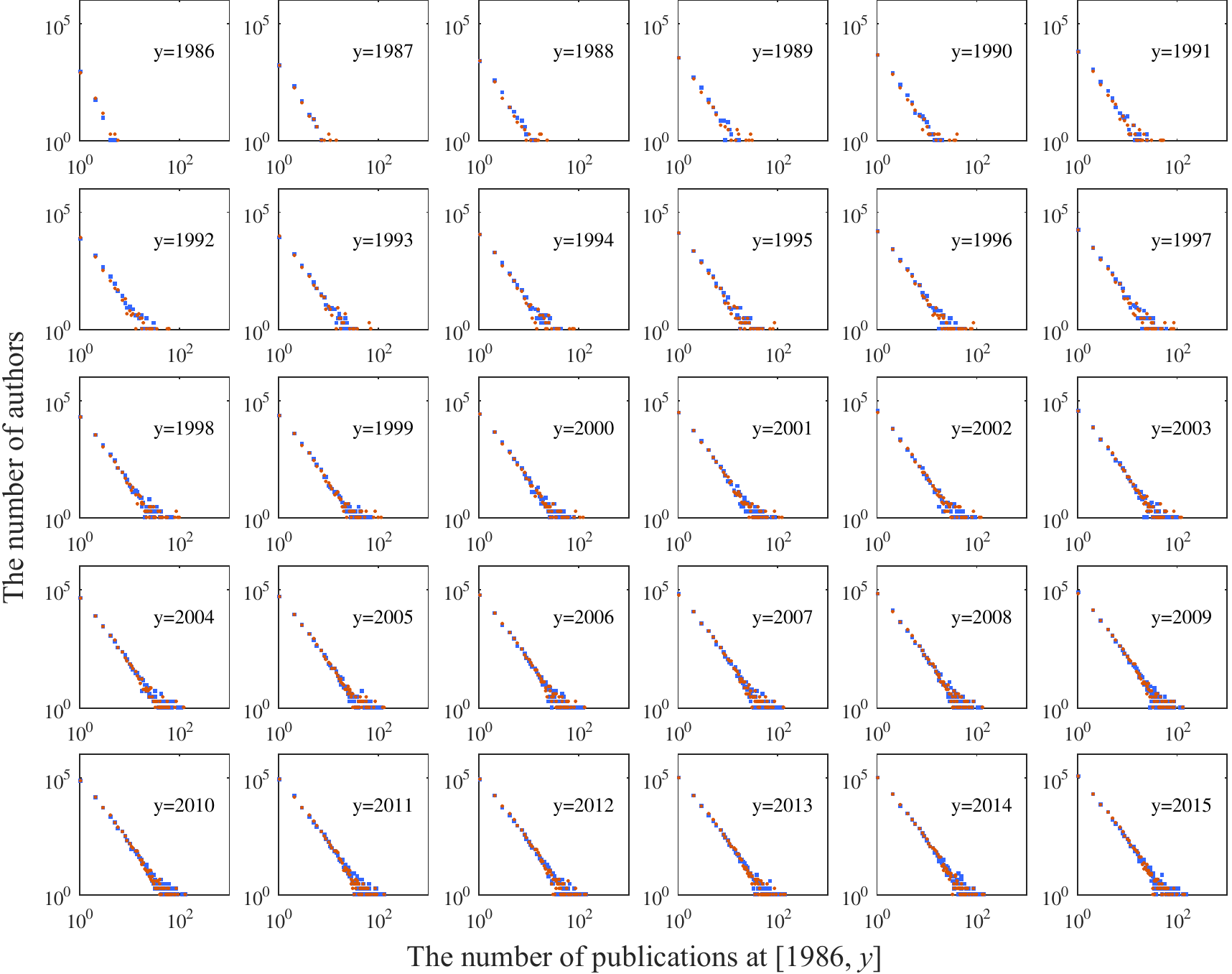}
\caption{    {\bf The distribution of the number of papers.  } Consider the
authors who  published papers  at $[1951,y]$, where $y=1986,...,2015$.
The panels  show the empirical  distribution (red circles) and predicted distribution  (blue squares)  for  the    authors  and their   papers published     at  the time interval $[1986 ,y]$.  }
  \label{hdegree}
\end{figure*}

\begin{figure*}[h]
\centering
\includegraphics[height=3.5   in,width=4.6     in,angle=0]{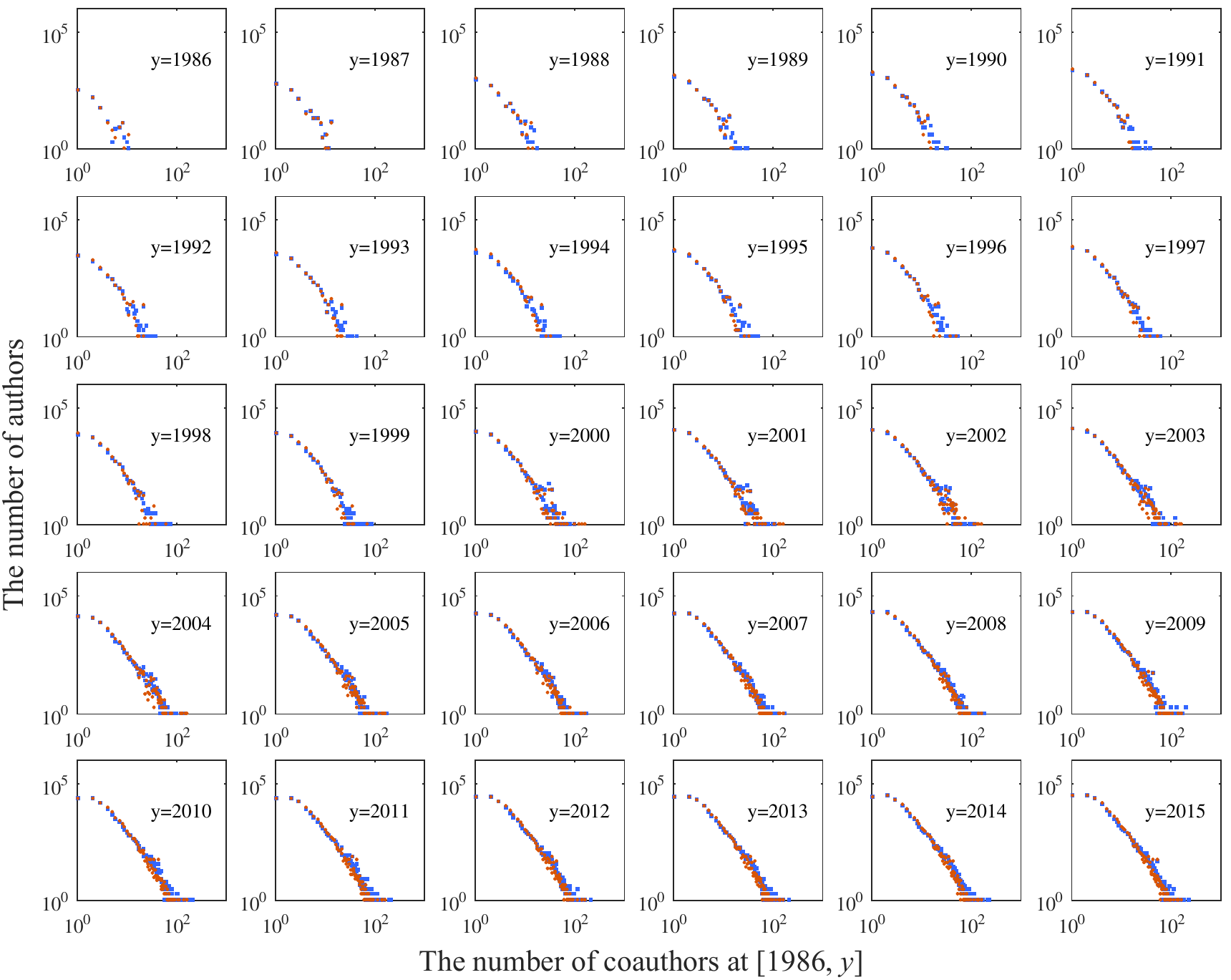}
\caption{    {\bf The     distribution of the number of coauthors.  }
 Consider the
authors who published papers   at $[1951,y]$, where $y=1986,...,2015$.
The panels  show the empirical  distribution (red circles) and predicted distribution  (blue squares)  for  the    authors  and their collaborations  at  the time interval $[1986 ,y]$.   }
  \label{degree}
\end{figure*}

We
compared the  empirical
degree and hyperdegree distributions   with the  those  predicted by
the proposed model.  Fig.~\ref{hdegree} and Fig.~\ref{degree} show that the model can  predict  the evolutionary trends of  the empirical degree and hyperdegree distributions. Moreover, at each year, the number of nodes with a given hyperdegree of the synthetic dataset. is also in close agreement
with that of the empirical dataset.


Fig.~\ref{deg_hd} has shown that
  the empirical dataset has the significantly positive correlation between
degree and hyperdegree,   which is measured by the Spearman rank-order
correlation coefficient.
 The model captures    this feature with a medium intensity  correlation coefficient close to that of the empirical  dataset, and well predicts the evolutionary trend of the correlation.

\subsection*{Clustering and degree assortativity}
\begin{figure*}[h]
\centering
\includegraphics[height=3.5   in,width=4.6     in,angle=0]{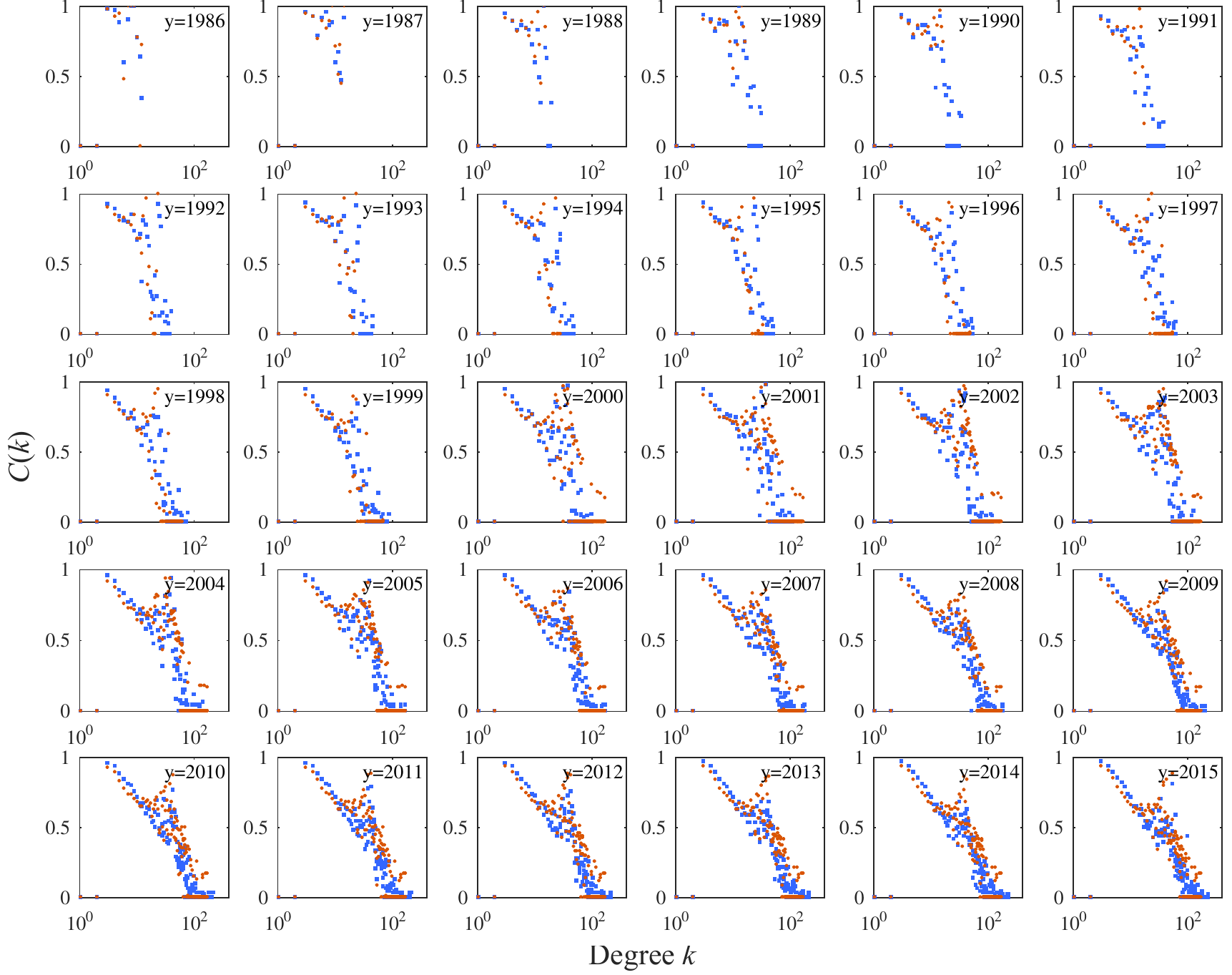}
\caption{    {\bf The     local clustering coefficient.  }  Consider the
authors who  published papers at $[1951,y]$, where $y=1986,...,2015$.
The panels show  the average local clustering coefficient of the authors with the same number of coauthors   at  the time interval $[1951 ,y]$ (red circles) and the predicted coefficient (blue squares). }
  \label{deg_ceo}
\end{figure*}

\begin{figure*}[h]
\centering
\includegraphics[height=3.5   in,width=4.6     in,angle=0]{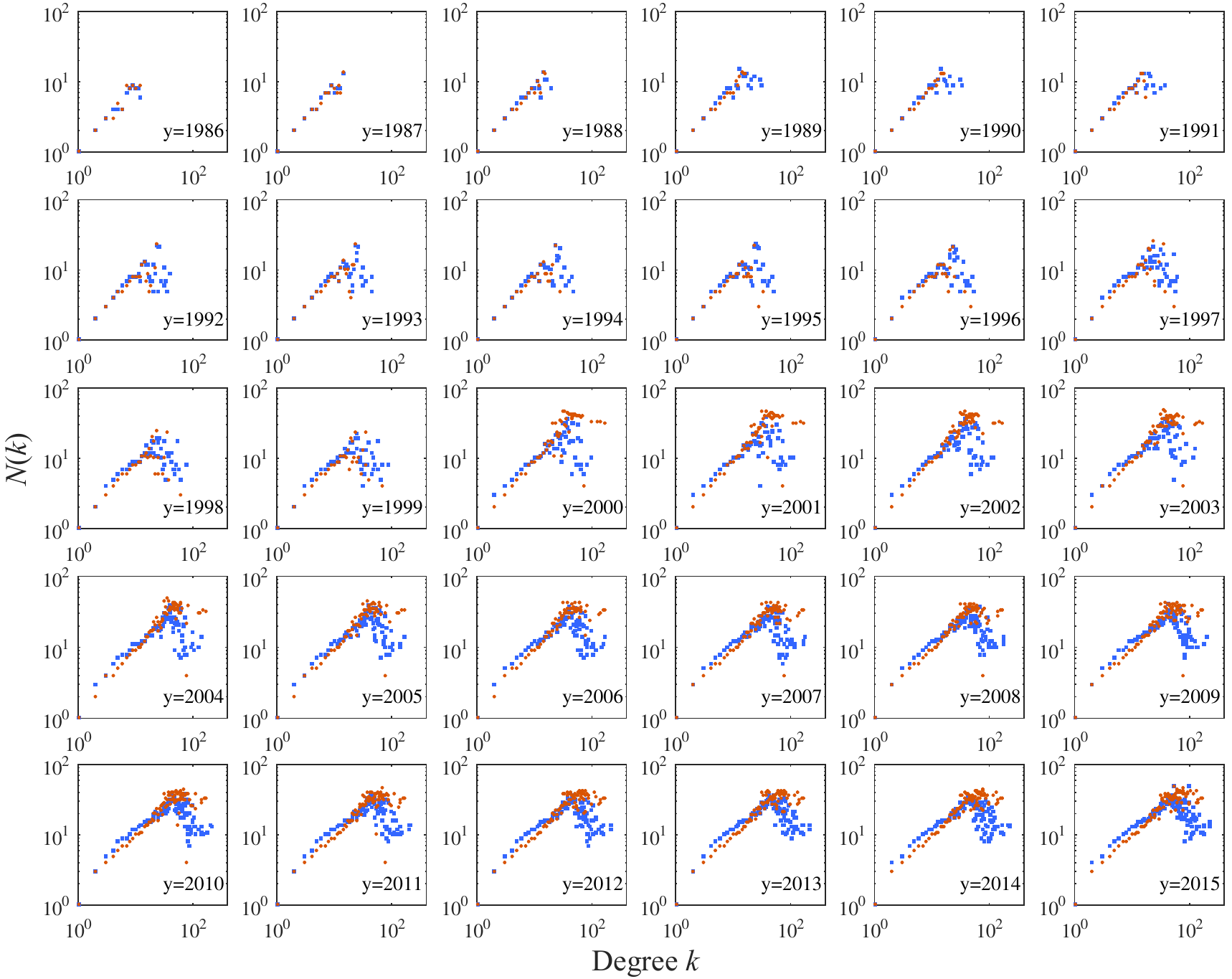}
\caption{    {\bf The  average number of the coauthors of neighbors.  } Consider the
authors who  published papers at $[1951,y]$, where $y=1986,...,2015$.
 The panels show  the average  number for the authors with the same number of coauthors    at  the time interval $[1951 ,y]$(red circles) and the predicted coefficient (blue squares).}
  \label{deg_nei}
\end{figure*}

Coauthorship networks are found to have two features: node clustering (neighbors of a
node probably connect to each other) and degree assortativity (the degree of a node positively correlates to the average degree of its neighbors), which are reflected through the
positive values of their global clustering coefficient and assortive coefficient (Table~\ref{tab1}).
Observing these features over degrees, we can find that those features differ from small degree nodes
to large degree nodes.
Denote the    average local clustering coefficient of $k$-degree nodes
by $C(k)$, and the average degree of $k$-degree nodes' neighbors by $N(k)$.
Fig.~\ref{deg_ceo} and Fig.~\ref{deg_nei}  show that
 the
model predicts  $C(k)$ and $N(k)$ of the empirical networks  well.

\begin{figure*}[h]
\centering
\includegraphics[height=3.5   in,width=4.6     in,angle=0]{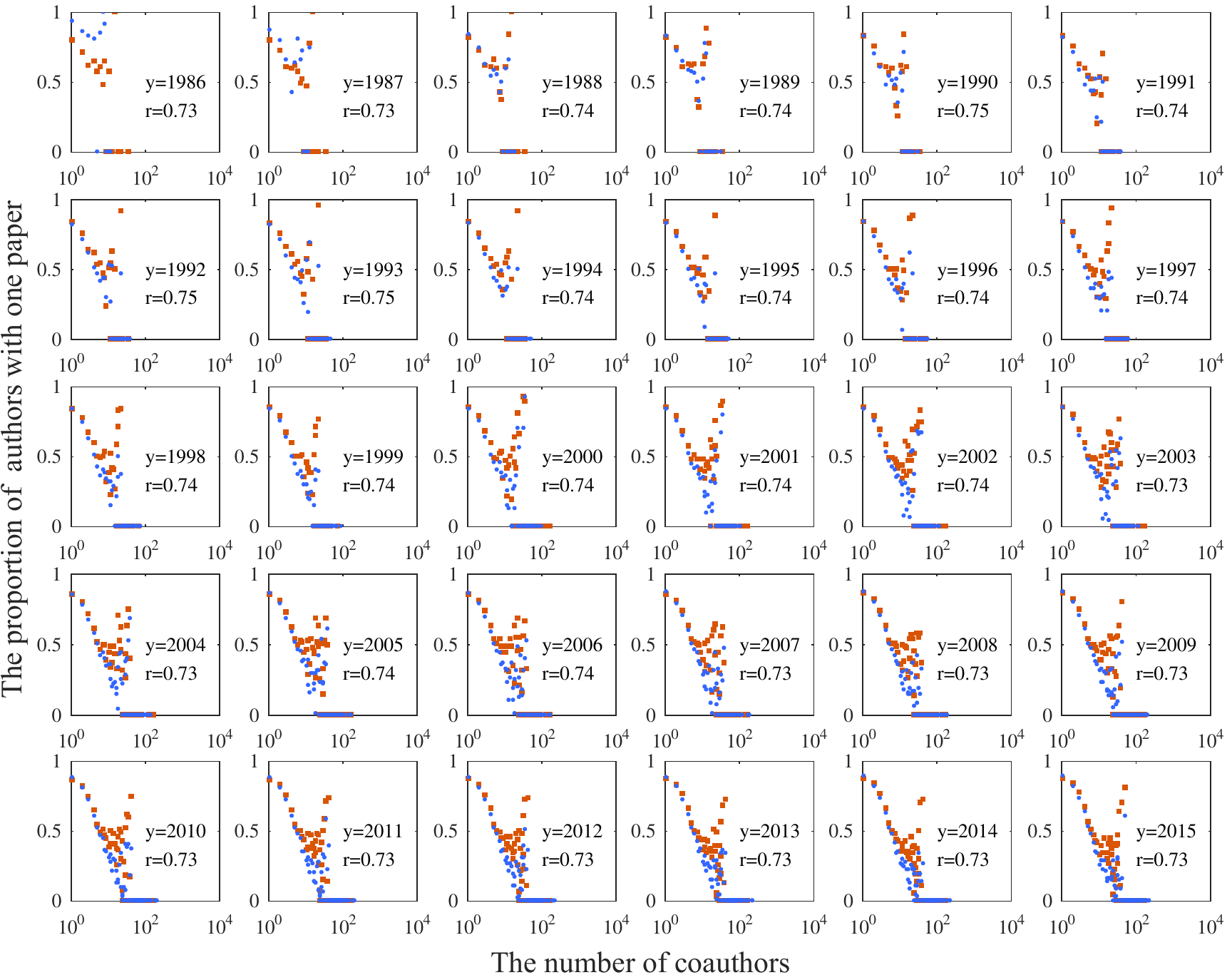}
\caption{    {\bf The proportion of the authors who publish  one paper.  } Consider the
authors who published papers at $[1951,y]$, where $y=1986,...,2015$.
The panels show
the proportion for   the authors with the same number of coauthors (red squares) and the predicted proportion (blue circles).
The   proportions   $r_t$ and $r_s$ are calculated based on all of the nodes for the empirical  and synthetic datasets.}
  \label{prop_hd_1}
\end{figure*}

Fig.~\ref{deg_nei}  also shows the model
gives a reasonable fit to the dichotomy of   $N(k)$ appeared in empirical data.
 The dichotomy can be derived based on three features of datasets:
a large fraction of nodes with hyperdegree one,
a large fraction of  hyperedges with a small size, and
the positive correlation between hyperdegree and degree.
The proposed model has those features.
The  sizes of  hyperedges,  as inputs of the model, are the same to the sizes of hyperedges of the empirical dataset.
Fig.~\ref{deg_hd} shows that the model captures
the positive correlation between hyperdegree and degree.
  Fig.~\ref{prop_hd_1} shows that the model  predicts  the proportion of $1$-hyperdegree nodes and the evolutionary trend
 of the proportion remarkably well.
Additionally,
this proportion taking high values  is   the reason
 for the  high clustering  of the empirical dataset, and thus  the   nodes in the synthetic dataset  are   also highly clustered.

   Consider  the nodes with a relatively small degree $k$.
 For the nodes with hyperdegree $1$ and degree $k$, their  neighbors     connect to  each other, and thus
their local clustering coefficient    is equal to $1$.
 Many of their  neighbors also
      belong to  one   hyperedge, and thus have    degree $k$.
      Therefore,
      regarding        the high proportion of  nodes with  one   hyperedge shown in Fig.~\ref{prop_hd_1},
   the value of $C(k)$ keeps high        over  relatively small   $k$, and the slope of $N(k)$ is    positive.

 Fig.~\ref{deg_hd} has shown that the positive correlation between degree and hyperdegree. That is,   a few  of  the   nodes  with a large degree $k$   would have a small hyperdegree, and many of they would also   have a large hyperdegree.
 Therefore, for the nodes with a large  degree $k$ and a large hyperdgree $s$, many of their  neighbors only belong to one  hyperedge, and thus
 their local clustering coefficient  decreases with the growth of    degree $k$.
Many of those neighbors
 have a small degree   equal to
the size of the corresponding hyperedge minus one, because    many hyperedges have a small size.
Therefore, the  slopes of   $C(k)$ and  $N(k)$ are  not   positive over large $k$.

\subsection*{Components and communities}

Table~\ref{tab1} shows the  networks extracted from the empirical dataset have giant component  and clear community structure,
  and the model captures these features.
A possible reason is shown as follows.
The nodes in the same hyperedge are very likely to belong to the same community.
We found that   the model cannot generate a giant component if $\varepsilon=0$.
Therefore, due to the small value of $\varepsilon$ used in the simulation, we could concluded that  the hyperedges are loosely connected in part caused  by   Algorithm~\ref{alg2}, or not connected at all.
 This
makes   the number of connections  within communities   significantly more than that between communities, and thus leads a  clear community structure.

 Algorithm~\ref{alg2}  controls the size of  giant component by
the parameter  $\varepsilon$. That is, $\varepsilon$ controls  the  phase transition from
 small components to
 the emergence of a giant component. It    simulates the phase transition from  isolated schools to invisible colleges.
If with a clear motivation,  we could  numerically investigate  the percolation transition. Percolation is
a central question in the study of random geometric graphs. This study does not involve it, only focuses on full-scale simulation.



\section*{Discussion and conclusions}
We proposed a distributed  model to simulate the  collaborations in the  dblp dataset for the researchers who  published papers in 1,304 journals and conference proceedings at $[1951, 2015]$. The  collaborations are expressed by a evolutionary hypergraph growing with time.
The model gives a full-scale simulation of a hypergraph that   grows from  5,099 nodes  to     149,285 nodes
with 106,821 hyperedges. From the perspective of coauthorship network,
  the model provides fine  fittings for the evolutionary trends of
     degree and hyperdegree distributions.
  Meanwhile, coauthorship patterns,  clustering  and degree assortativity as well as their evolutionary trend   predicted by the model are also in close agreement with those of the empirical dataset.
The assembly mechanism of the model    provides an example of how the evolution of collaborations can be derived by two possible mechanisms the cumulative advantage and  the
individual  strategies based on
 maximizing the   benefit-cost ratio, and how
 the  network complexity of  collaborations emerge
in the evolution.

  The parameters of the model   give  flexibility to fit  other empirical   datasets.
 Therefore, it  has the potential to be a null model for studying  social affiliation networks with heterogeneous multinary relationship.
 However,  the  mathematical formulae underlying the predicted degree and hyperdegree distributions  are derived  based on  the orders of the fitting polynomials of
the trend of annual number of papers and the cumulative number of authors.
The trend of the dblp dataset from 2016 cannot be well fitted by the polynomials used in this study. Therefore, we only simulated the collaborations up to 2015. Additionally,
after 2015, we found some authors
published more than one hundred  papers in a year, which  also  cannot be predicted by our model.
 It means   that the assembly mechanism
 of the model should be modified   when applying it to other empirical studies.
In addition, some typical  mechanisms of   cooperation should be  considered, such as voluntary participation, group selection, and social diversity~\cite{Perc2008,Perc2010}.
 And the factors  of geography\cite{Hoekman-Frenken2010} and discipline or interdiscipline\cite{Rijnsoever-Hessels2011} also need  to be considered.
%





The
phenomena emerged in human behaviors are
usually quite complex. Yet,
little is known about the mechanisms governing the evolution   of
publication productivity and collaboration behaviors of
researchers, whilst our model
  renders evolution  trajectories   relatively predictable on average by Lotka's law and a cooperative game.
  However,  we should note that  the  difference on the evolutionary of degree assortativity coefficient between the empirical dataset and the synthetic dataset is still unknown to us.
    Analyzing massive data  to track scientific careers would   help to
  solve it, and will  improve our understanding of
    how collaborations evolve.
 Some learning models, e.g.,   recurrent neural networks,   can give
    good short- or long-term predictions for individuals in terms of the number of papers and that of coauthors\cite{Xie2020JInfor,Xie2020TCSS}; thus  we are trying to find a way to integrate their advantages.



 \section*{Acknowledgments} The author thanks   Professor Jinying Su in the
National University of Defense Technology for her helpful comments
and feedback. This work is supported by the   National Natural   Science Foundation of China (Grant No. 61773020) and  National Education Science Foundation of China (Grant No. DIA180383).



\end{document}